\documentclass[]{spie}

\usepackage[utf8]{inputenc}
\usepackage{amsmath,amsfonts,amssymb}
\usepackage{graphicx}
\usepackage[colorlinks=true, allcolors=blue]{hyperref}
\usepackage{upgreek}
\usepackage{epstopdf}
\usepackage{epsfig}
\usepackage{units}

\usepackage{acronym}
\usepackage[caption=false]{subfig}

\acrodef{AO}[AO]{adaptive optics}
\acrodef{CCD}{charged-coupled device}
\acrodef{DM}[DM]{deformable mirror}
\acrodef{FLAO}{First Light AO}
\acrodef{ISYS}[ISYS]{Institute for System Dynamics Stuttgart}
\acrodef{KIT}[KIT]{Karlsruhe Institute for Technology}
\acrodef{KOOL}[KOOL]{Koenigstuhl Observatory Opto-Mechatronics Laboratory}
\acrodef{LBT}[LBT]{Large Binocular Telescope}
\acrodef{LSW}[LSW]{Landessternwarte}
\acrodef{MCF}[MCF]{multi-core fiber}
\acrodef{MFD}[MFD]{mode-field diameter}
\acrodef{MLA}[MLA]{micro-lens array}
\acrodef{MM}[MM]{multi-mode}
\acrodef{MMF}[MMF]{multi-mode fiber}
\acrodef{MPIA}[MPIA]{Max Plank Institute for Astronomy}
\acrodef{NA}[NA]{numerical aperture}
\acrodef{NCP}[NCP]{non-common path}
\acrodef{NAIR}[NAIR]{Novel Astronomical Instrumentation based on photonic light Reformating}
\acrodef{NIR}[NIR]{near-infrared}
\acrodef{PSF}[PSF]{point-spread function}
\acrodef{PSD}[PSD]{power spectral density}
\acrodef{SM}[SM]{single-mode}
\acrodef{SMF}[SMF]{single-mode fiber}
\acrodef{WFS}[WFS]{wavefront sensor}
\acrodef{ExAO}[ExAO]{extreme adaptive optics}

\newcommand{\mum}{\upmu \mathrm{m}}

\title{Micro-lens array as tip-tilt sensor for single-mode fiber coupling}

\author[a]{Philipp Hottinger}
\author[a]{Robert J. Harris}
\author[b,c,d]{Philipp-Immanuel Dietrich}
\author[b,c]{Matthias Blaicher}
\author[e]{Martin Glück}
\author[f]{Andrew Bechter}
\author[f]{Jonathan Crass}
\author[g]{Jörg-Uwe Pott}
\author[b,c,d]{Christian Koos}
\author[e]{Oliver Sawodny}
\author[a]{Andreas Quirrenbach}

\affil[a]{Landessternwarte~(LSW), Zentrum für Astronomie der Universität Heidelberg, Königstuhl~12, 69117~Heidelberg, Germany}
\affil[b]{Institute of Microstructure Technology~(IMT), Karlsruhe Institute of Technology~(KIT), Hermann-von-Helmholtz-Platz~1, 76344~Eggenstein-Leopoldshafen, Germany}
\affil[c]{Institute of Photonics and Quantum Electronics~(IPQ), Karlsruhe Institute of Technology~(KIT), Engesserstr.~5, 76131~Karlsruhe, Germany}
\affil[d]{Vanguard Photonics GmbH, Hermann-von-Helmholtz-Platz 1, 76344~Eggenstein-Leopoldshafen, Germany}
\affil[e]{Institute for System Dynamics, University of Stuttgart, Waldburgstr. 19, 70563 Stuttgart, Germany}
\affil[f]{Department of Physics, University of Notre Dame, 225~Nieuwland Science Hall, Notre Dame, IN~46556, USA}
\affil[g]{Max-Planck-Institute for Astronomy, Königstuhl~17, 69117~Heidelberg, Germany}

\authorinfo{Further author information: (Send correspondence to P.H.)\\P.H.: E-mail: phottinger@lsw.uni-heidelberg.de, Telephone: +49 6221 54 1727}

\begin{document}
\maketitle

\begin{abstract}

We introduce a design for a tip-tilt sensor with integrated \acl{SMF} coupling for use with the front-end prototype of the iLocater spectrograph at the \acl{LBT} to detect vibrations that occur within the optical train.
This sensor is made up of a \acl{MLA} printed on top of a fiber bundle consisting of a central \acl{SMF} and six surrounding \aclp{MMF}.
The design in based on a previous prototype that utilized a \acl{MCF} with seven \aclp{SMF} \cite{Dietrich2017}. With this updated design, we are able to achieve a better sensing throughput.
We report on the modeled performance: if the beam is perfectly aligned, $69\%$ light is coupled into the central \acl{SMF} feeding the scientific instrument. When the beam is not aligned, some of the light will be coupled into the outer sensing fibers, providing the position of the beam for tip-tilt correction.
For this design we show that there is a linear response in the sensing fibers when the beam is subject to tip-tilt movement.
Furthermore we introduce an adaptive optics testbed, which we call the Koenigstuhl Observatory Opto-mechatronics Laboratory (KOOL), this testbed currently simulates vibrations at the \acl{LBT}, and in collaboration we have extended it to allow \acl{SMF} coupling tests.

\end{abstract}


\section{Introduction}
\label{sec:introduction}
\acresetall 

For many years the image quality of ground based telescopes was limited by the atmosphere, known as the seeing limit. However, recent advances in modern adaptive optics (AO) systems are allowing 8-10~m class telescopes to achieve better imaging quality, leading to new and exciting discoveries. In particular \ac{ExAO} can allow diffraction limited imaging in certain circumstances. Examples of these systems include FLAO at the Large Binocular Telescope (LBT, 2x8.4~m)\cite{Esposito2011}, GPI at the Gemini South Observatory (8.2~m)\cite{Macintosh2014}, SCExAO at the Subaru Telescope (8.2~m)\cite{Jovanovic2015} and SPHERE at the Very Large Telescope (VLT, 8.2~m)\cite{Beuzit2008}.

Conventional fiber-fed spectrographs use \acp{MMF} as the different modes of the telescopes \ac{PSF} need to be propagated.
Yet, improved developments in \ac{AO} open up the new possibility to use spectrographs fed by \ac{SM} fibers, instead of larger \acp{MMF}. Due to the smaller entrance aperture, or slit, these spectrographs can be reduced in size, reducing stability constraints and are also free of conventional modal noise \cite{Crepp2016}. Several attempts have been made to couple \acp{SMF} to these large telescopes, but the coupling efficiency is strongly affected by the quality of initial fiber alignment, as well as beam drifts and higher-frequency tip-tilt motions due to telescope or instrument mechanics and vibrations.

Conventional tip-tilt sensing solutions include imaging of the \ac{PSF} on a quad-cell detector\cite{Esposito1997}, imaging a pinhole mirror, using the telescopes \ac{AO} system, or accelerometer based disturbance feed-forward control\cite{Gluck2017}. Yet, these approaches can suffer from fundamental limitations that limit their accuracy. These include \ac{NCP} vibrations, limited dynamical range, low response speed and additional throughput losses.
In this work we introduce a tip-tilt sensor consisting of a \ac{MLA} printed on top of a fiber bundle that overcomes many of these limitations. The design is optimized to perform both tip-tilt sensing using \acp{MMF} and simultaneously couple light into a \ac{SMF} to feed the spectrograph.

This concept is based on a prototype device introduced by Ref.~\citenum{Dietrich2017} but uses \acp{MMF} for sensing to improve sensitivity.
The \ac{MLA} will be printed on top of the fiber bundle by in-situ two-proton lithography to produce these free-form lenses and achieve high alignment precision\cite{Dietrich2018}.
The design is optimized to be installed in the front-end prototype of the iLocater spectrograph at the \ac{LBT} to increase \ac{SMF} coupling efficiency.


In Sec.~\ref{sec:design_considerations} we introduce the iLocater spectrograph, its optical properties and a short analysis of its tip-tilt vibration challenges.
Sec.~\ref{sec:final_design} describes the preliminary design of the tip-tilt sensor taking into account the requirements of the telescope and instrument including its modeled performance and manufacturing plans.
Sec.~\ref{sec:KOOL} introduces the \ac{AO} testbed, \ac{KOOL} where we are performing tests for \ac{SMF} coupling and tip-tilt sensing.
This is followed by Sec.~\ref{sec:discussion}, which outlines advantages and a comparison to a first prototype described by Ref.~\citenum{Dietrich2017}, and Sec.~\ref{sec:conclusion} summarizes and highlights future work.

\section{Design considerations}
\label{sec:design_considerations}

We are developing the tip-tilt sensor to be integrated with the \ac{SMF} coupling front-end prototype for the iLocater spectrograph. For this it is essential to understand the instrument and its requirements.

This section gives an overview over the iLocater spectrograph (Sec.~\ref{sec:ilocater}), its requirements for \ac{SMF} coupling (Sec.~\ref{sec:SMF-coupling}) and the observed vibrations (Sec.~\ref{sec:vibrations}).

\subsection{iLocater spectrograph}
\label{sec:ilocater}

iLocater is a high resolution spectrograph for the \ac{LBT} \cite{Crepp2016}. A fiber injection system feeds the cross-dispersed Echelle spectrograph that operates in the YJ-bands (0.97-1.27~$\mum$). The instrument will deliver a high spectral resolving power ($\mathrm{R}\sim150,000$).

Unlike most conventional high resolution spectrograph, it will be fed by \acp{SMF}.
For this, light from both 8.4~m diameter telescopes of the \ac{LBT} is corrected by the LBTI \ac{AO} system \cite{Hinz2012} and then coupled into the \acp{SMF}.
The spectrograph accommodates three spectroscopic input channels: one for each telescope and one for the wavelength calibration source fed by a Fabry-P\'erot etalon calibration system for radial velocity precision below 10 cm/s \cite{Stuermer2017}.

Using \acp{SMF} to feed the spectrograph yields several advantages over traditional \acp{MMF}.
\acp{SMF} feature a smaller output aperture and a lower \ac{NA}, leading to a compact instrument design while achieving high spectral resolution. The iLocater spectrograph will have a footprint of 50~cm squared \cite{Crepp2016}. When compared with other spectrographs on  8~m class telescopes (e.g. Ref.~\citenum{Pepe:2010}) this is small, reducing cost and increasing stability.
Furthermore, \acp{SMF} are free of conventional modal noise\cite{Halverson2015}, though recent studies have shown that polarization may also cause noise in a high resolution spectrograph \cite{Halverson2015}.

\subsection{Single mode fiber coupling}
\label{sec:SMF-coupling}

While \acp{SMF} offer many advantages, efficient light coupling from a telescope into the fibers can prove to be a challenging task and several requirements have to be considered.
Within a \ac{SMF}, light will propagate in only one mode, the fundamental mode with a near-Gaussian intensity profile. Its width (\ac{MFD}) and the relative refractive indicies of the core and cladding govern the \ac{NA} of the light exiting or entering the fiber as approximated by a Gaussian beam.

The beam from the telescope on the other hand is most similar to an Airy pattern due to diffraction at the primary mirror (though this is slightly altered by secondary obscuration, spiders etc.). The \ac{NA} at the fiber coupling plane is set by the telescope optics. This diffraction pattern and the \ac{NA} need to be closely matched with the fundamental mode of the \ac{SMF} to achieve maximum coupling efficiency. This can be calculated by the overlap integral of the incoming beam and the accepted near-Gaussian intensity profile.
 As there is still a fundamental mismatch between the Airy intensity distribution and the fundamental mode, the theoretical maximum coupling efficiency is $\sim80\%$\cite{Shaklan1988} without efforts such as pupil apodization attempting to overcome this limitation\cite{Guyon2003}.

\begin{figure}
	\begin{center}
		\begin{tabular}{c c}
			\subfloat[]{
        \includegraphics[width=.4\textwidth]{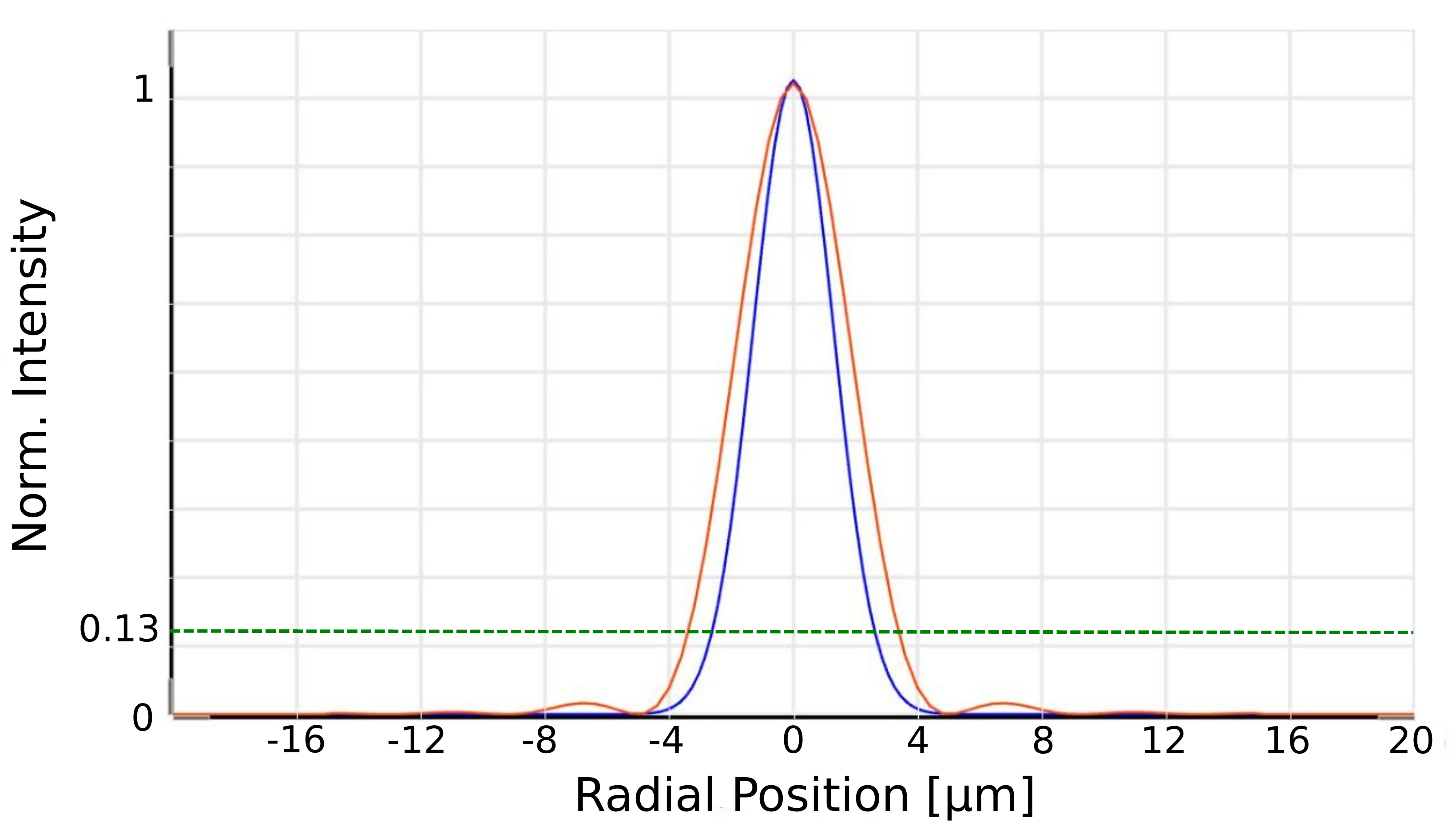}
        \label{fig:SMF-profiles}
      } &
			\subfloat[]{
        \includegraphics[width=.4\textwidth]{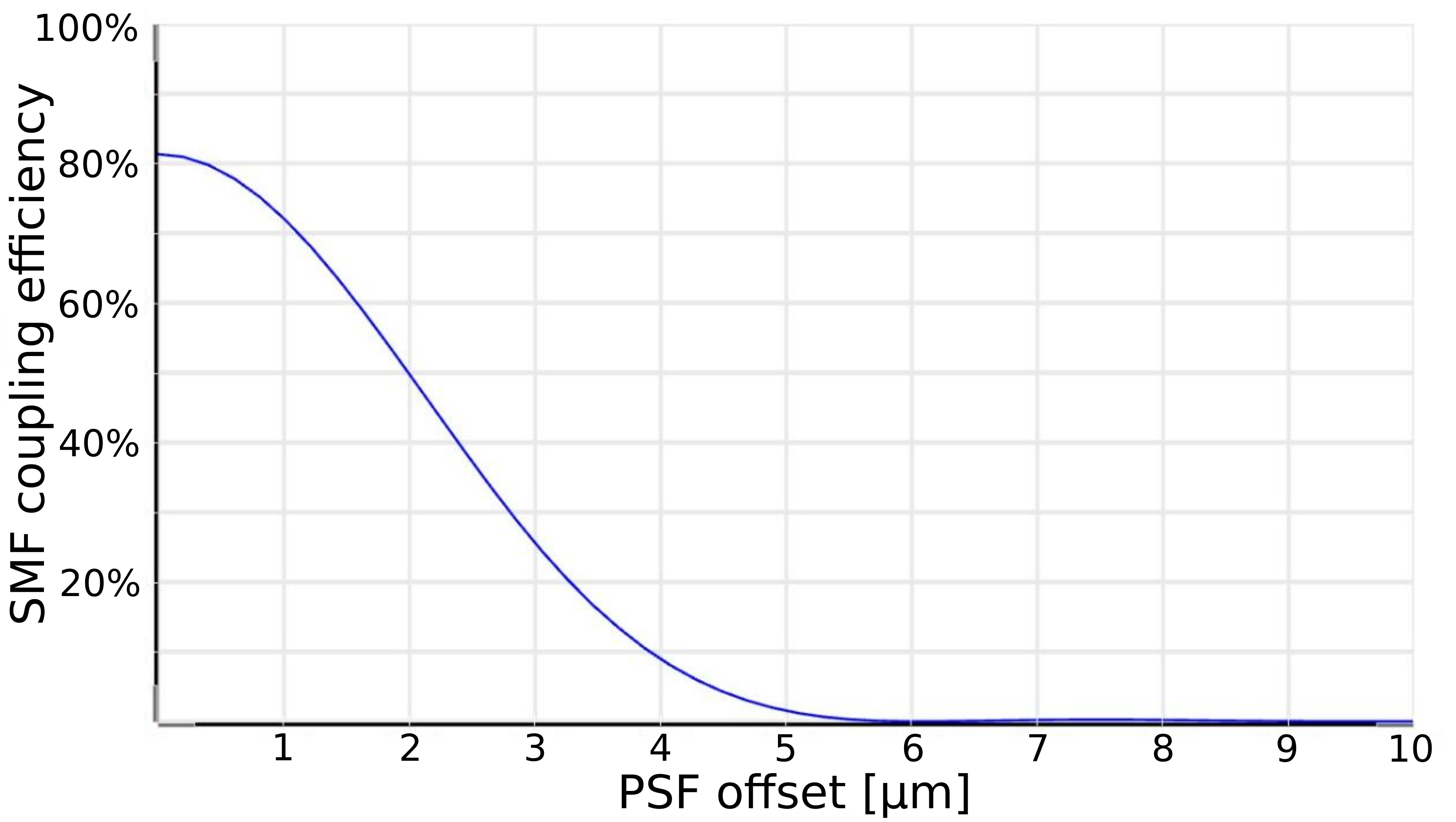}
        \label{fig:SMF-offset}
      }
		\end{tabular}
	\end{center}
	\caption{
  Simulated \acf{SMF} coupling for a \acf{MFD} of $5.8~\mum$ ($1/\mathrm{e}^2$ intensity).
  Image \textbf{(a)} shows the cross section of the intensity profile for both the fundamental mode of a \acl{SMF} (blue) and a diffraction limited \acl{PSF} optimized for maximum coupling efficiency into that \ac{SMF} (red). The green dashed line indicates the $1/e^2$-intensity (13\% of the maximum).
  Image \textbf{(b)} shows the coupling efficiency if the incoming beam is gradually misaligned with respect to the fiber by shifting the centroid of the \acf{PSF}, resulting in a rapid decrease in coupling efficiency with position.
  }
	\label{fig:SMF-coupling}
\end{figure}

Fig.~\ref{fig:SMF-profiles} shows the intensity profile of both the Airy pattern of an idealized telescope (red) and the fundamental mode of the fiber (blue) as modeled with the the optical design software Zemax (see Sec.~\ref{sec:modeling}), optimized for a \ac{SMF} with \ac{MFD} of $5.8~\mum$ ($1/\mathrm{e}^2$ intensity). Note, as the diffraction limit is usually measured as distance between maximum and first minimum, the size of the diffraction limited \ac{PSF} is defined somewhat larger ($\sim10~\mum$ diameter between first minima).
If the incoming beam is not perfectly aligned to the fiber, the coupling efficiency is further reduced. This is plotted in Fig.~\ref{fig:SMF-offset} for a beam gradually misaligned from the \ac{SMF} by decentering its centroid position, showing a rapid decrease in coupling efficiency. With a beam displacement of $2.8~\mum$ corresponding to the mode-field radius of the \ac{SMF}, the coupling efficiency is reduced to $\sim26\%$.
This shows the precision that is necessary to efficiently couple light from the telescope into the \ac{SMF} and the large impact a slight misalignment has on throughput.


As the \ac{SMF} chosen for the iLocater instrument has a \ac{MFD} of $5.8\mum$ ($1/\mathrm{e}^2$-diameter) the incoming beam must be optimized for coupling to this.
For this, the F/15 beam of the \ac{LBT} and the diffraction pattern of 60~mas at a wavelength of $1~\mum$ (diameter to the first minima)
are re-imaged by the iLocater front-end\cite{Bechter2015}.

\subsection{Tip-tilt vibrations}
\label{sec:vibrations}

It is well known that the \ac{LBT}, as most telescopes, suffers from vibrations\cite{Brix2008}. This effect is particularity pronounced when \ac{ExAO} is used, as the vibration relative to \ac{PSF} size is larger.
When testing \ac{SMF} injection for iLocater with a prototype front-end bench in 2015, coupling of up to 25\% was shown \cite{Bechter2016}. This was lower than the theoretical maximum of $\sim80\%$\cite{Shaklan1988} due to several reasons.
Extensive tests have shown that most of this is accounted to vibrations throughout the telescope and the fiber injection bench. These vibrations cause tip-tilt wavefront aberrations leading to a movement of the \ac{PSF} on the focal plane and thus preventing more efficient fiber coupling.

The following data is based on images taken by an ANDOR Zyla 5.5 camera set up at the imaging arm of the iLocater front-end working in a wavelength range between 700~nm and 970~nm\cite{Bechter2016}.
The exposure time is $\sim$1~ms and the sampling frequency is $\sim$125~Hz.

\begin{figure}
	\begin{center}
		\includegraphics[width=0.5\textwidth]{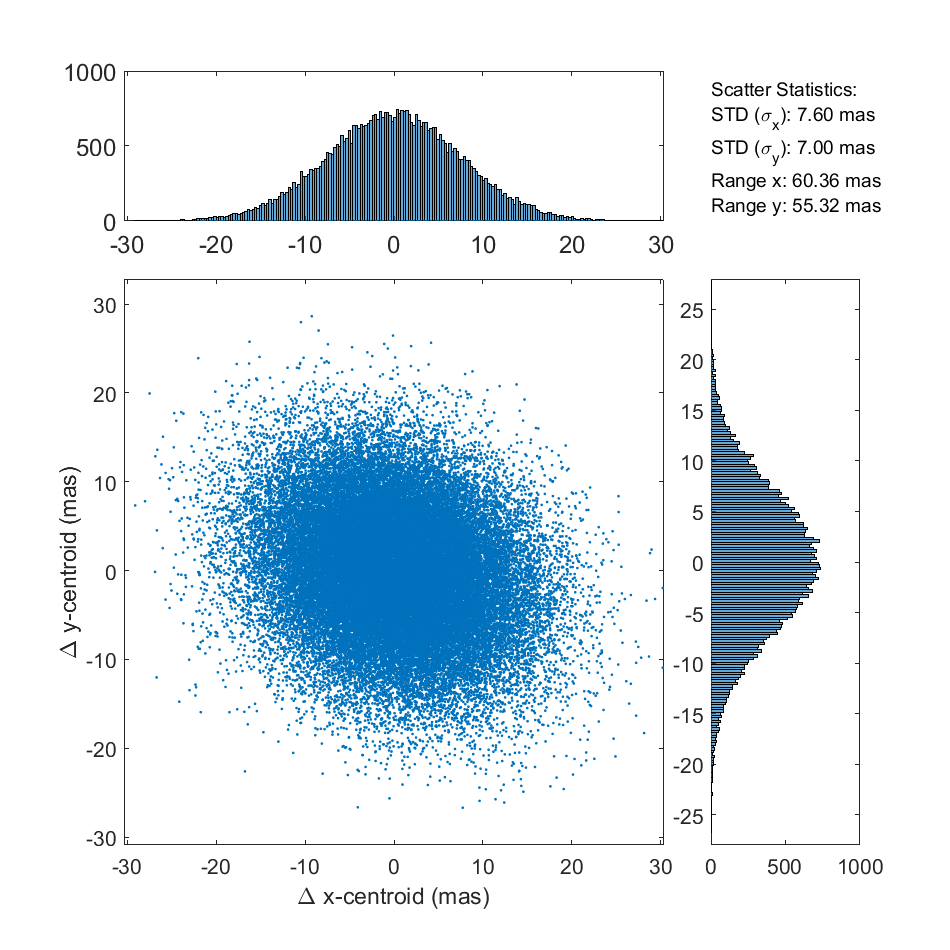}
	\end{center}
	\caption{
  Scatter plot of the location of the \ac{PSF} centroid for each exposure and the corresponding distributions. The wavelength range of 700~nm to 970~nm was imaged by the iLocater front-end prototype as described in Ref.~\citenum{Bechter2016}. The diameter of the diffraction limited \ac{PSF} is $\sim$60~mas.
  }
  \label{fig:psf_move}
\end{figure}

\begin{figure}
	\begin{center}
		\begin{tabular}{c c}
			\subfloat[]{\includegraphics[height=5.5cm]{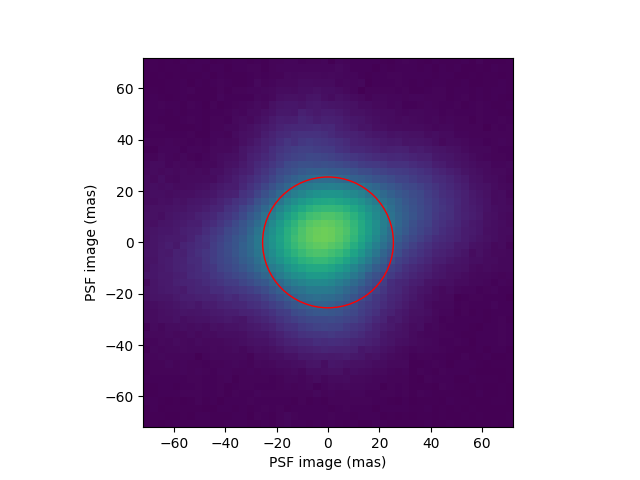}} &
			\subfloat[]{\includegraphics[height=5.5cm]{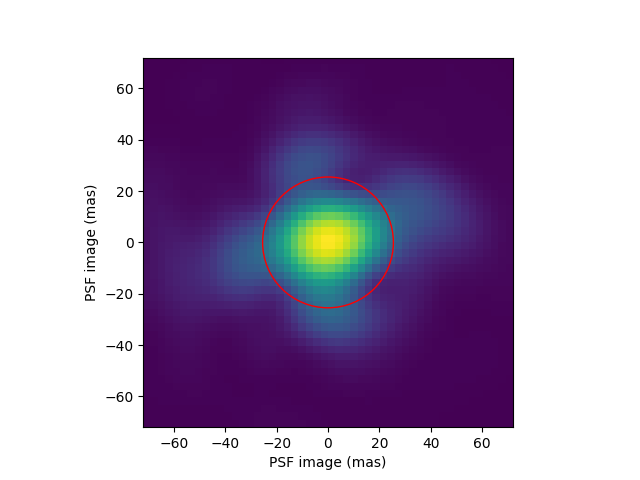}}
		\end{tabular}
	\end{center}
	\caption{
  Averaged Point Spread Function from the \ac{LBT}, with \textbf{(a)}  no tip-tilt compensation and \textbf{(b)} with tip-tilt compensation calculated by holding the center of a fitted Gaussian constant. Both images have logarithmic intensity scaling and are normalized to the highest pixel value. The red circle denotes the Airy disk diameter at 850 nm, indicating that with vibrational compensation the coupling will be higher. It can also be seen from b) that aberrations in the system need to be compensated for.
  }
  \label{fig:psf_comparison}
\end{figure}

Fig.~\ref{fig:psf_move} shows the position of the \ac{PSF} center for each exposure, both as scatter plot on the image plane and as distribution for each axis.
The \ac{PSF} is misaligned with respect to its mean position by up to 60~mas, corresponding to more that one diffraction limited \ac{PSF} diameter, leading to loss due to inefficient fiber coupling as outlined in Sec.~\ref{sec:SMF-coupling}.
To optimize throughput, these tip-tilt vibrations need to be sensed and corrected.

Fig.~\ref{fig:psf_comparison} shows stacked images of the \ac{PSF}. Image (a) shows the sum of all exposures simulating a longer exposure time. Image (b) also shows the sum of all exposures but with the image shifted in such a way that the center of a fitted Gaussian is held constant, effectively simulating a perfect tip-tilt correction. The red circle indicates the diffraction limit at the working wavelength ($\sim$850~nm).
These images illustrate that despite working near the diffraction limited regime, tip-tilt vibrations can smear out the \ac{PSF} and lead to a seeing limited result complicating \ac{SMF} coupling.
Note that even for the tip-tilt corrected image (b), higher order aberrations are visible. These are most likely static \ac{NCP} aberrations that also need to be corrected.

\section{Preliminary design}
\label{sec:final_design}

\begin{figure}
	\centering
	\includegraphics[width=0.5\textwidth]{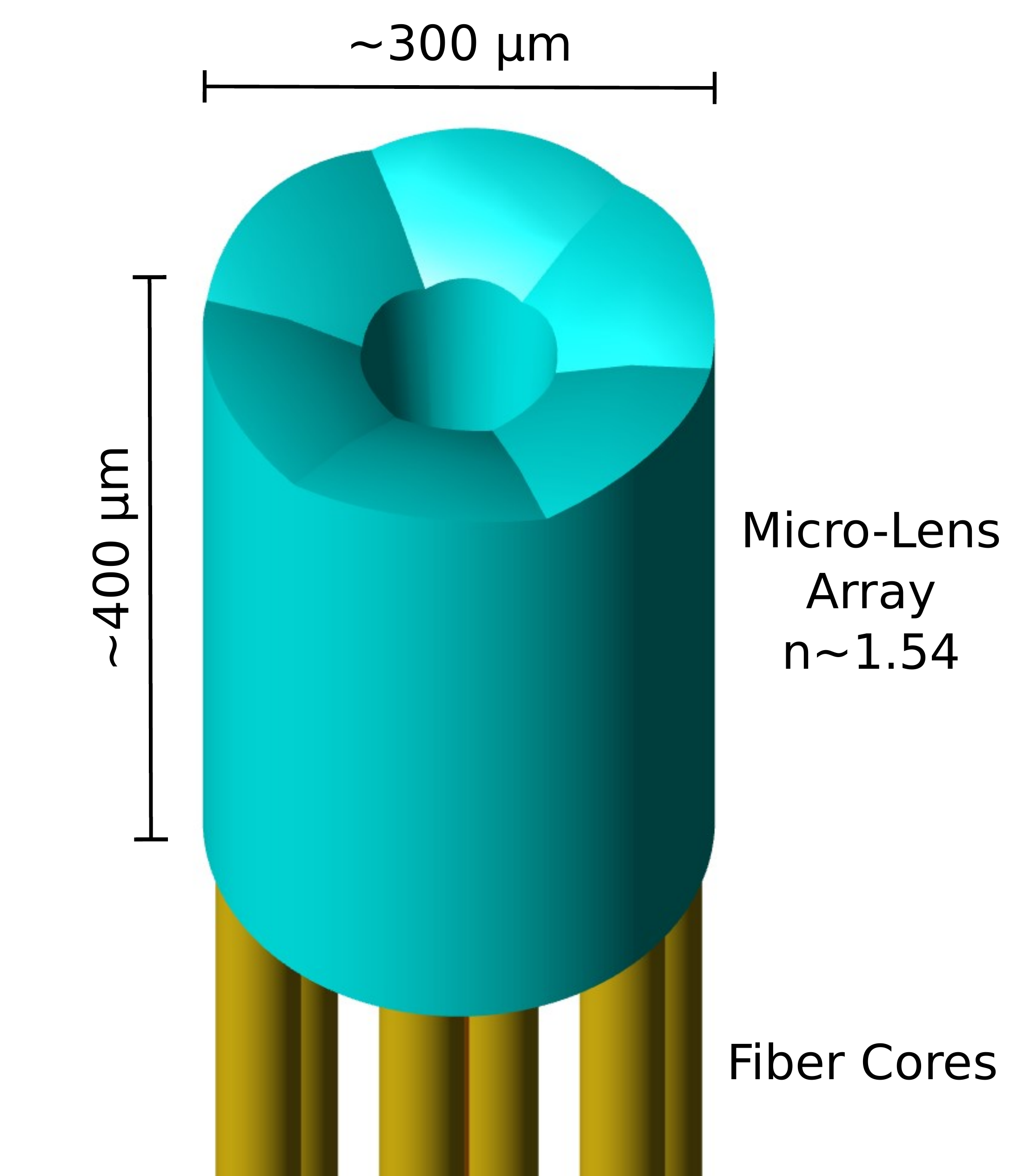}
	\caption{
	Design model of a \acf*{MLA} tip-tilt sensor for the front-end prototype of the iLocater spectrograph.
	The \ac{MLA} is printed on top of a fiber bundle made of a central \acf*{SMF} (not visible) and six surrounding \acfp*{MMF} (only fiber cores shown).
	While the central optical path to the \ac{SMF} is unobscured, off-centered light will be refracted by the \ac{MLA} to be coupled into the surrounding sensing \acp{MMF}.
  The pitch between the fiber cores is $125~\mum$. Diameter and height of the \ac{MLA} are $300~\mum$ and $400~\mum$, respectively
  }
\label{fig:MLA_model}
\end{figure}

To optimize coupling into the \ac{SMF} and therefore throughput, the tip-tilt motions of the incoming telescope beam as described in Sec.~\ref{sec:vibrations} need to be sensed and corrected for.
There are some conventional methods that have been used to do such tip-tilt sensing, yet these can have some fundamental limitations (see Sec.~\ref{sec:introduction} and Sec.~\ref{sec:discussion}).

In this work we present an innovative method for tip-tilt sensing, based upon the work in Ref.~\citenum{Dietrich2017}.
This sensor consists of a \acf{MLA} printed on top of a fiber bundle made up of a central \ac{SMF} (\ac{MFD} of $5.8~\mum$) and six surrounding \acp{MMF} (core diameter of $50~\mum$, \ac{NA} of 0.22) arranged in a hexagonal array with a pitch of $125~\mum$.
The \ac{MLA} itself covers just all fiber cores, has a diameter of $300~\mum$ and a height of $\sim400~\mum$.
The central \ac{SMF} serves as the fiber that will feed the spectrograph, while the tip-tilt sensing signal will be provided by coupling into the surrounding \acp{MMF}.
Fig.~\ref{fig:MLA_model} shows a 3D model of this system.

\begin{figure}
	\centering
	\begin{tabular}{cccc@{}}
		\subfloat[aligned]{\includegraphics[width=0.2\textwidth,trim={150 0 150 0},clip]{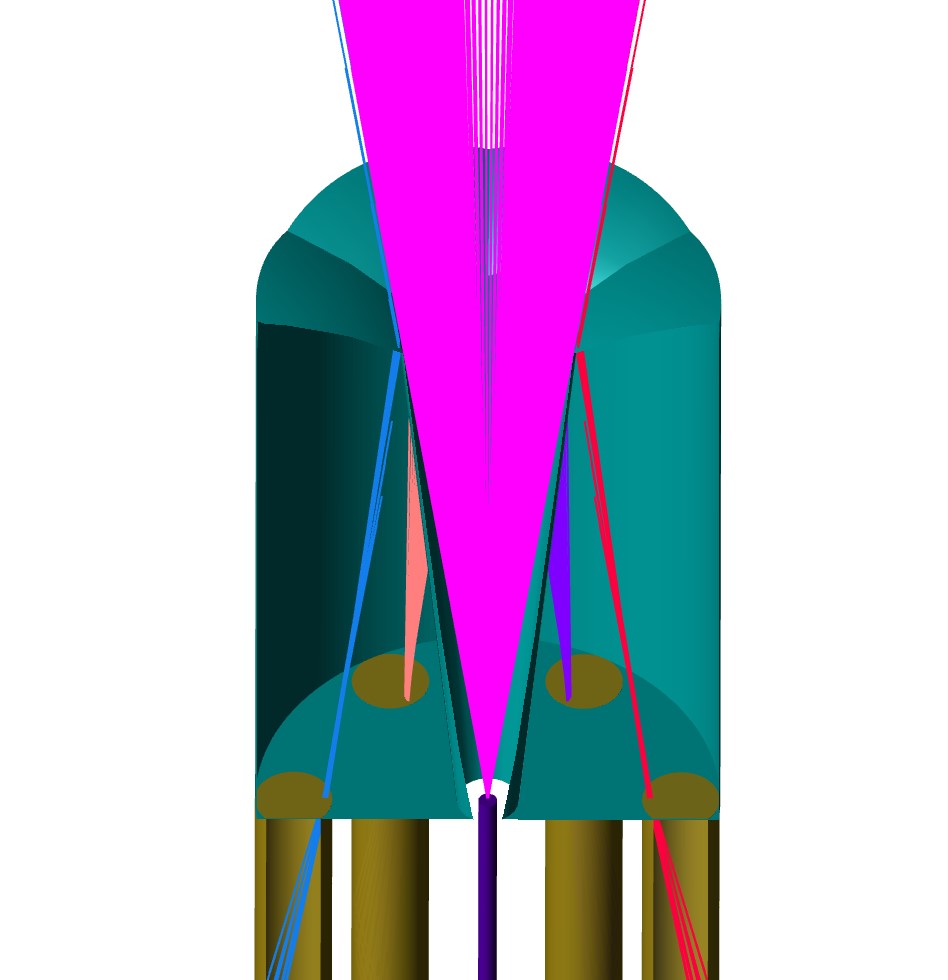} \label{fig:MLA_center}} &
		\subfloat[$2~\mum$]{\includegraphics[width=0.2\textwidth,trim={150 0 150 0},clip]{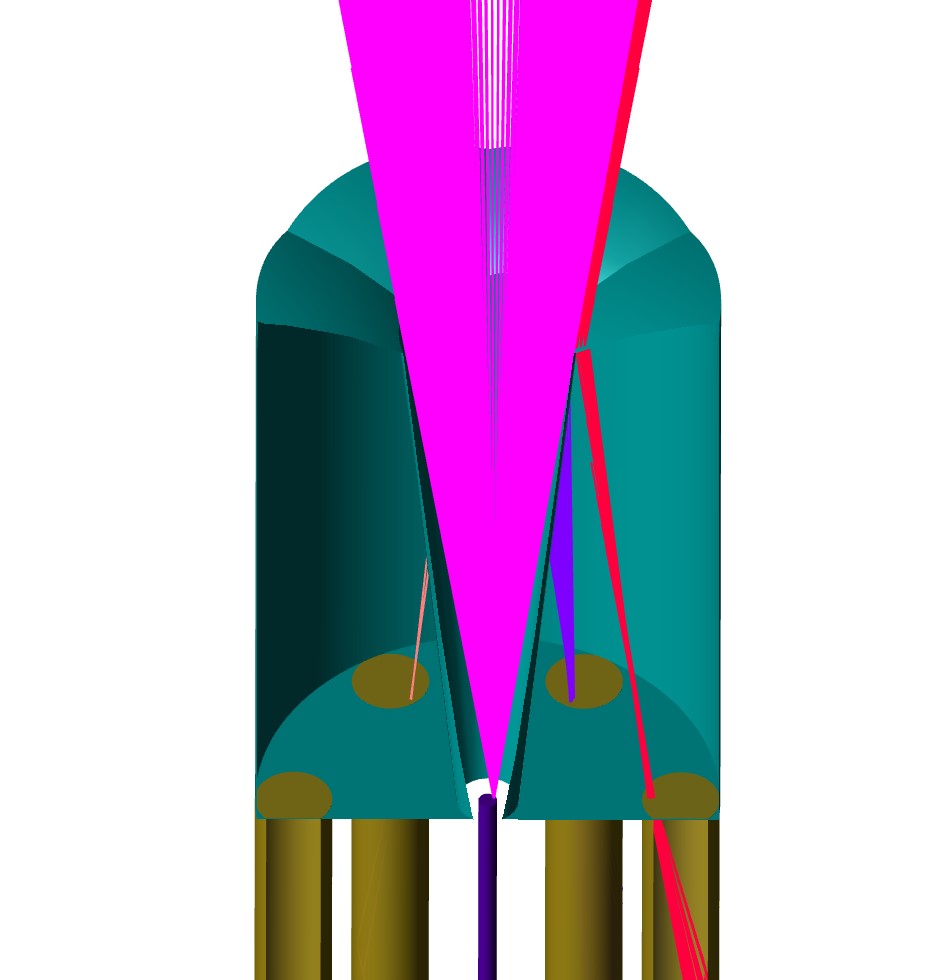} \label{fig:MLA_mis1}} &
		\subfloat[$10~\mum$]{\includegraphics[width=0.2\textwidth,trim={150 0 150 0},clip]{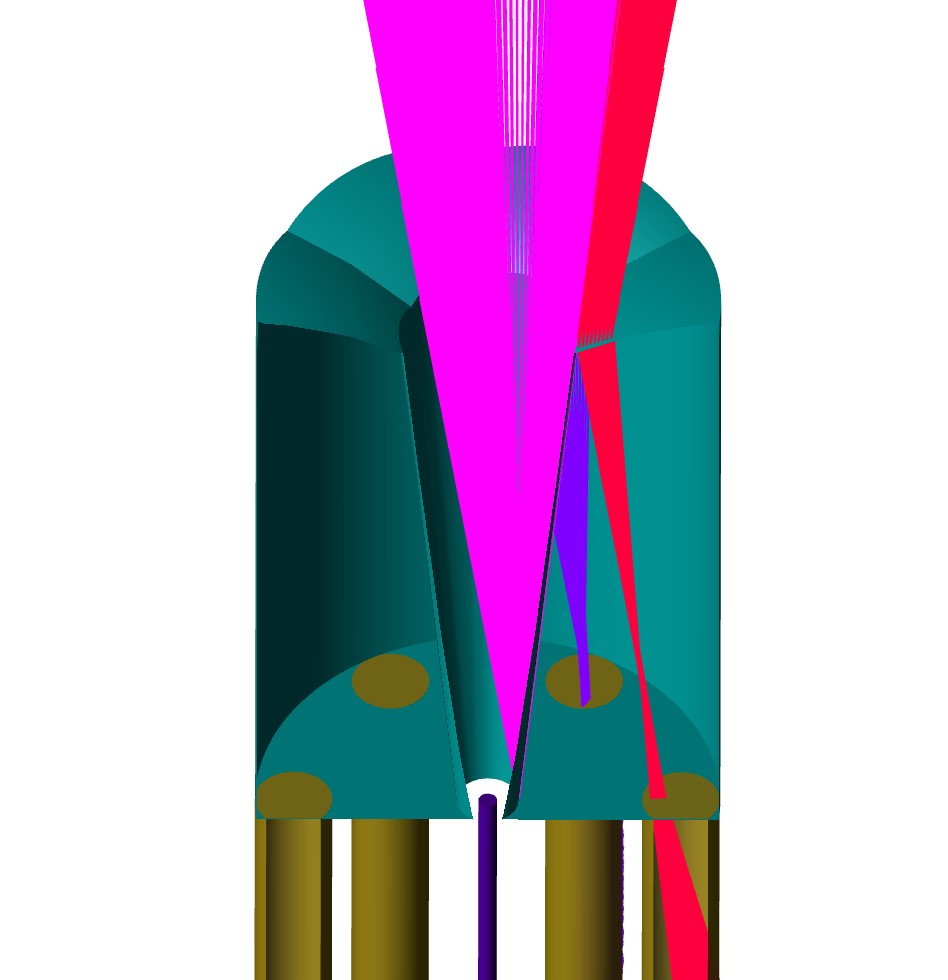} \label{fig:MLA_mis2}} &
		\subfloat[$50~\mum$]{\includegraphics[width=0.2\textwidth,trim={150 0 150 0},clip]{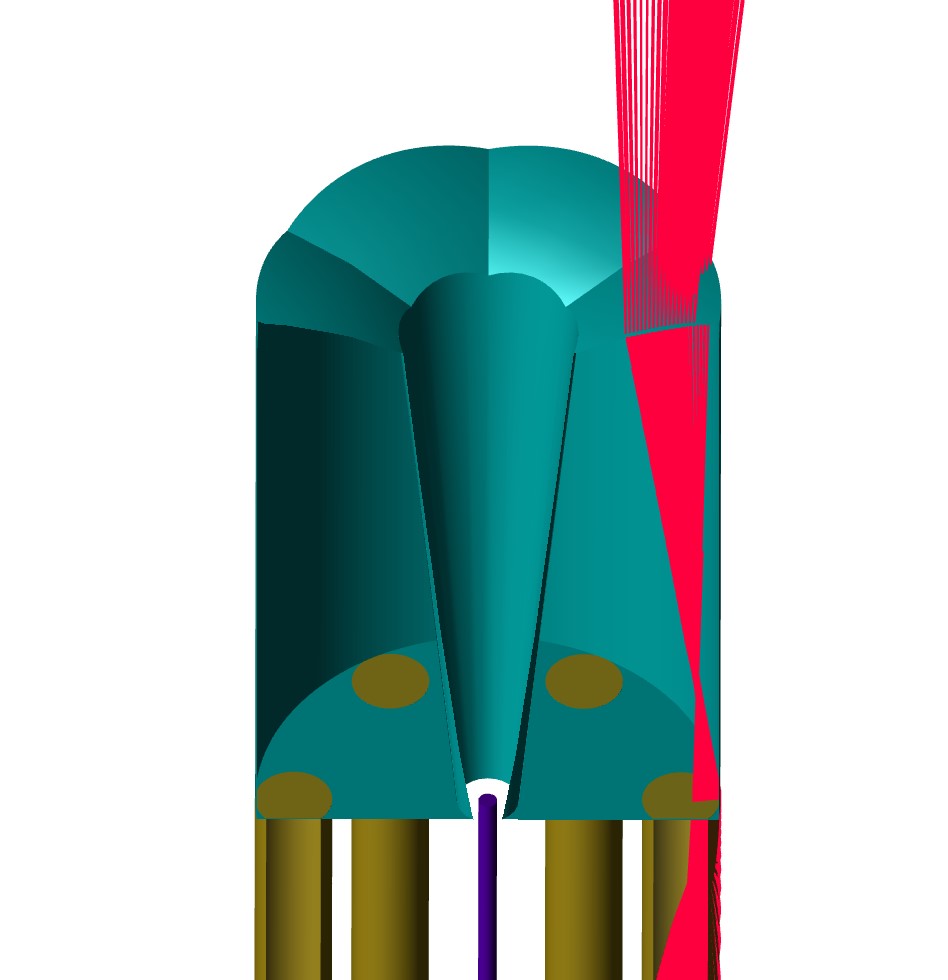} \label{fig:MLA_mis3}}
	\end{tabular}
	\caption{
	Cross section of the \acf*{MLA} tip-tilt sensor with incoming beam illustrating the working principle of the device. The color of the rays indicate which fiber the light is coupled into: the central \acf*{SMF} feeding the spectrograph (purple, number of rays does not correspond to coupling efficiency) and the outer sensing \acfp*{MMF} (red, violet, orange, blue). Rays that are refracted by the \ac*{MLA} and are not coupled into any fiber are not illustrated.
  If the beam is perfectly aligned \textbf{(a)}, most of the light is coupled into the central \ac*{SMF} with some of the rest coupled into the outer sensing \acp*{MMF}.
  As the beam is decentered by $2~\mum$ \textbf{(b)}, coupling efficiency into the \ac*{SMF} decreases but is still significant while the sensing signal in the outer \acp*{MMF} increases.
  For large misalignments of $10~\mum$ \textbf{(c)} and $50~\mum$ \textbf{(d)} no light is coupled into the central \ac*{SMF} but successively more light is coupled into the sensing \acp*{MMF} illustrating the wide dynamical range the tip-tilt sensor covers.
	}
	\label{fig:MLA_crosssection}
\end{figure}

Fig.~\ref{fig:MLA_crosssection} illustrates the working principles of the tip-tilt sensor. It shows the cross-section of the \ac{MLA} model for four different incoming beam positions indicated by the captions.
Image (a) shows a beam perfectly aligned with the \ac{SMF}. In this case 69\% of the light is coupled into this fiber. Only some light will be refracted evenly to the six surrounding \acp{MMF}.
Image (b) shows a slightly misaligned beam, where most of the light is still coupled into the central \ac{SMF} but more light is coupled into the outer sensing fibers in direction of the decentered beam. In this regime, real time tip-tilt sensing and correction will take place.
Image (c) and (d) show even more misaligned beams with hardly any coupling into the central \ac{SMF} but successively more light coupling into the sensing fiber.
As hardly any light is coupled into the \ac{SMF} if the beam is misaligned this much, this regime is not favorable for real-time correction. Nevertheless, this illustrates the wide dynamical range of the sensor, extending its use to initial fiber and target alignment.

The following section will outline its modeled performance, design considerations (Sec.~\ref{sec:modeling}), correction method (Sec.~\ref{sec:correction}) and manufacturing plans (Sec.~\ref{sec:manufacturing}).

\subsection{Modeling}
\label{sec:modeling}

To optimize the \ac{MLA} and to model its performance, extensive tests were conducted. These were performed with the optical design software \textit{Zemax OpticStudio} \cite{zemax}.
\textit{Zemax OpticStudio} uses a ray tracing algorithm for lens design. This is usually not suitable for simulating coupling into \acp{SMF} as they to not take into account wave properties of the incoming beam.
The wave nature of the beam is of fundamental importance as it forms the diffraction pattern described by wave optics that is responsible for matching the incoming telescope beam to the fundamental mode of the \ac{SMF} (see Sec.~\ref{sec:SMF-coupling}).
Because of this, the \textit{Physical Optics Propagation} capabilities of \textit{Zemax OpticStudio} where used which take into account both Gaussian and wave optics for \ac{SMF} coupling while the ray tracing capabilities were used for \ac{MMF} coupling.

\subsubsection{Sensing performance}
\label{sec:performance}

\begin{figure}
 \centering
 \includegraphics[width=0.6\textwidth]{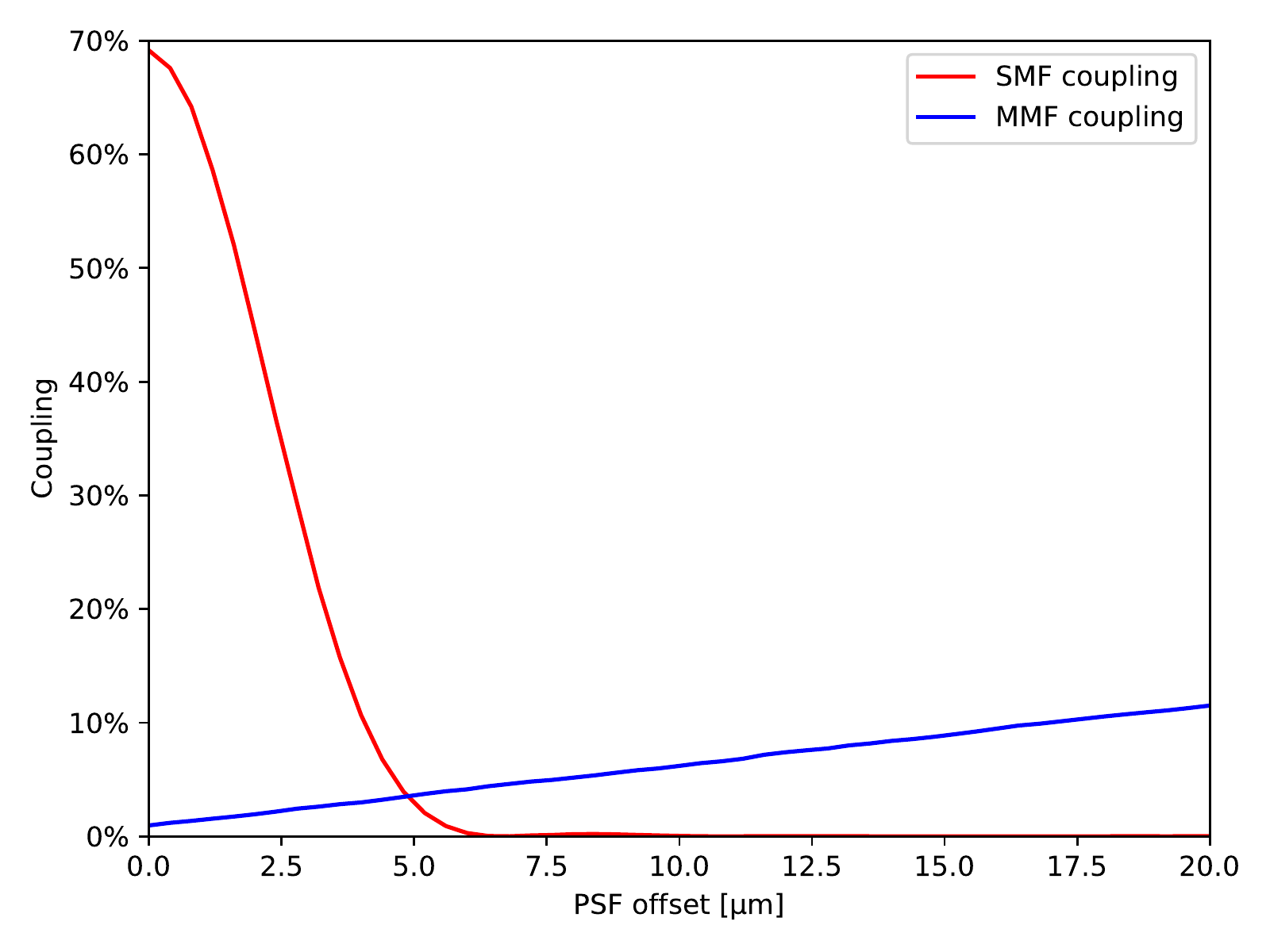}
\caption{
Modeled coupling into the \ac*{SMF} (red line) and into the sensing \ac*{MMF} (blue line) for an incoming beam moving from a perfectly aligned position along an axis to this sensing fiber.
}
\label{fig:MLA_coupling}
\end{figure}

The modeled coupling efficiencies are shown in Fig.~\ref{fig:MLA_coupling}, with the red line showing the fiber coupling efficiency for the central \ac{SMF} and the blue line showing the coupling efficiency for one of the sensing \ac{MMF} as the incoming beam is decentered.
The coupling efficiency for the central \ac{SMF} decreases rapidly as the beam is decentered.
When the beam is misaligned by $5~\mum$, corresponding to the $1/e^2$-\ac{MFD} of the modeled fiber, total coupling has already decreased to $<5\%$, which is similar to a standalone \ac{SMF} (see Sec.~\ref{sec:SMF-coupling}).
The coupling into the sensing \ac{MMF} increases linearly starting at $\sim1\%$ for an aligned beam to roughly $20\%$ for an offset of $20\ \mum$.
As the signal for this offset is linear, this design should prove to be easy to integrate with the existing tip-tilt mirror in the iLocater front-end prototype.

\subsubsection{Off-axis performance}
\label{sec:MLA_axis}

\begin{figure}
	\begin{center}
		\begin{tabular}{c c}
			\subfloat[]{
        \includegraphics[height=6cm,trim={0 0 0 300},clip] {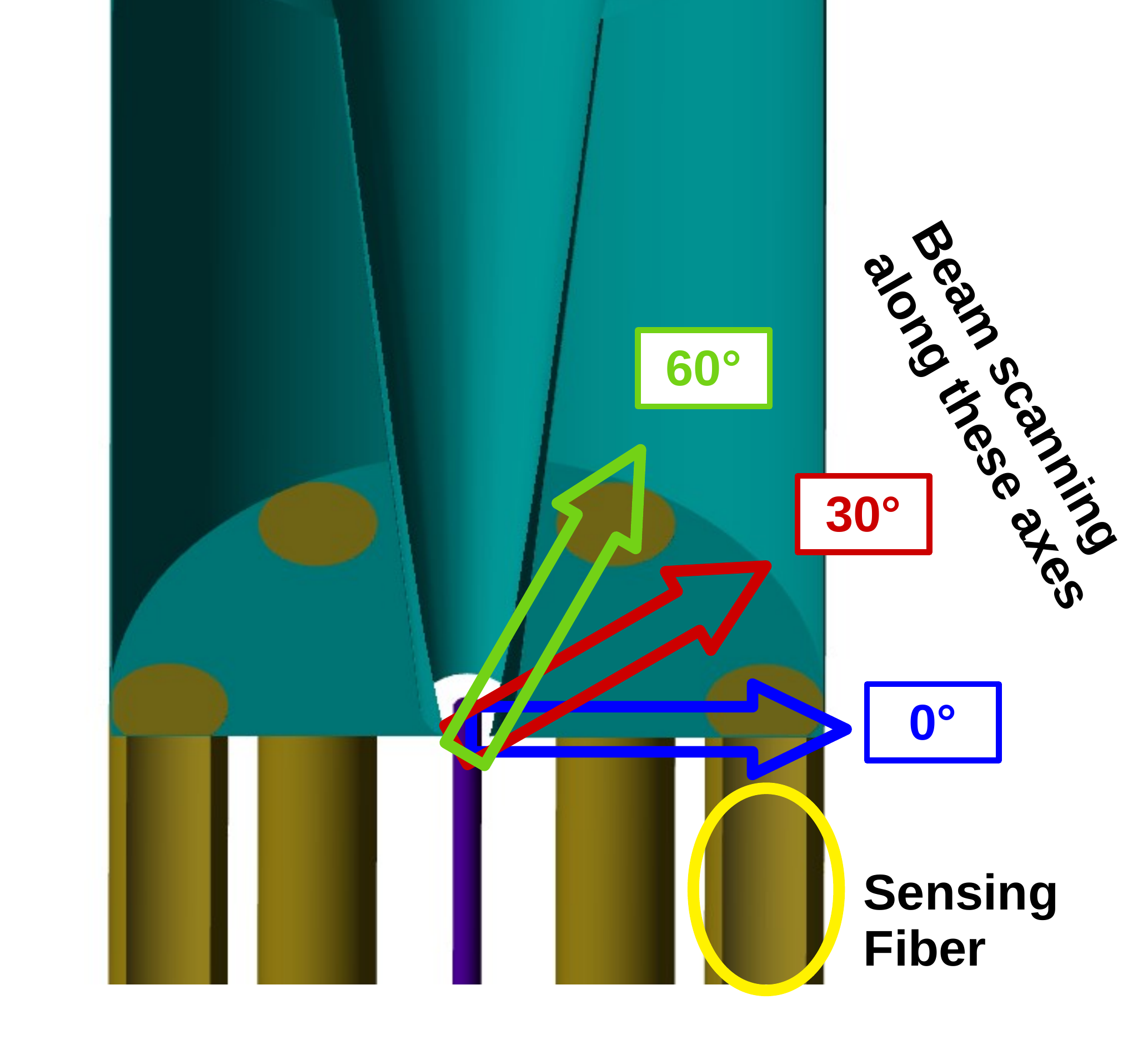}
        \label{fig:coupling_60deg_model}
        } &
			\subfloat[]{
        \includegraphics[height=6cm] {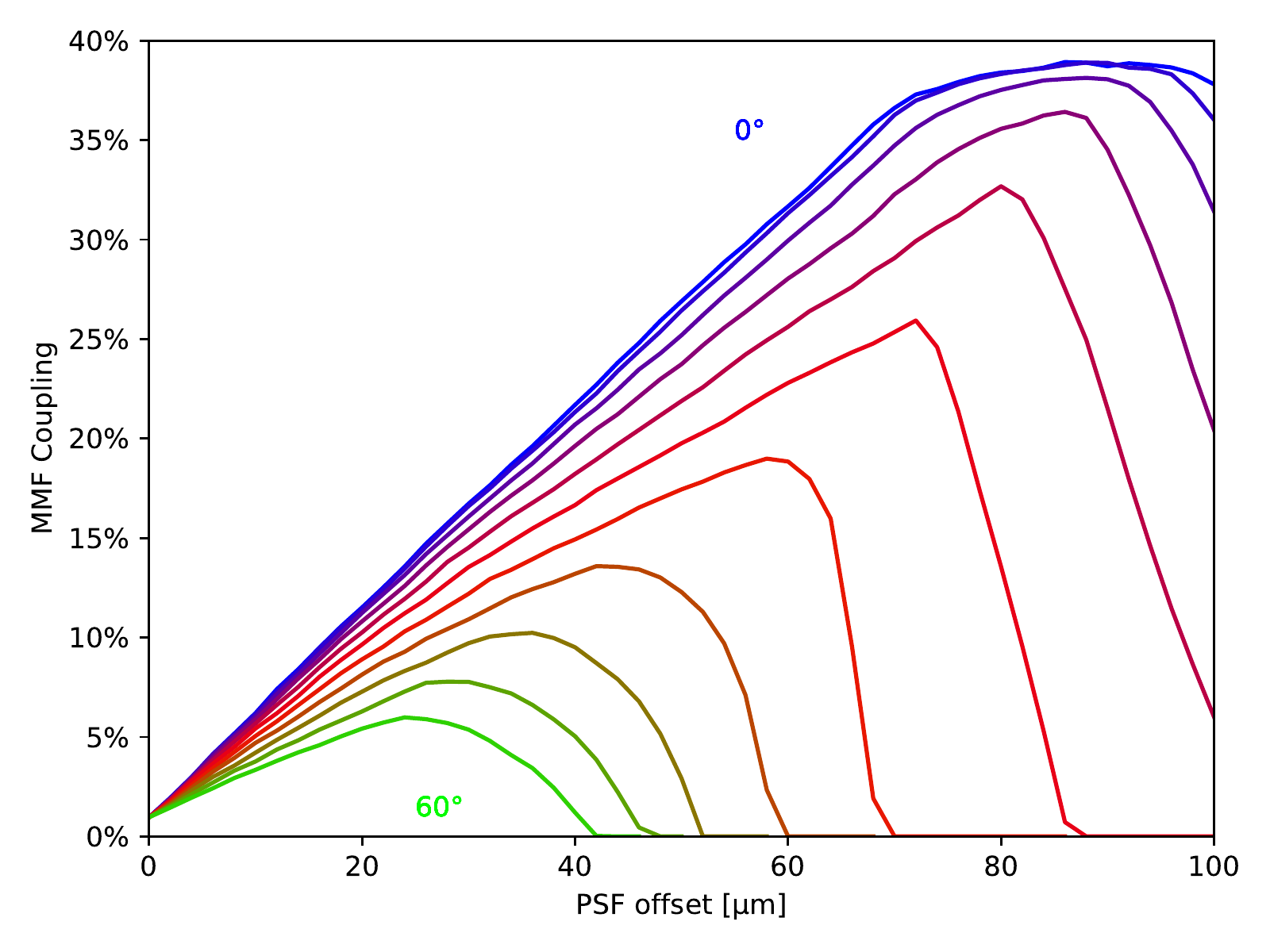}
        \label{fig:coupling_60deg_plot}
        }
		\end{tabular}
	\end{center}
	\caption{
  Off-axis performance of the \ac*{MLA} tip-tilt sensor.
  Modeled coupling for one sensing \ac*{MMF} for an incoming beam moving from an aligned position outwards.
  This is illustrated in the image \textbf{(a)}: The rightmost fiber is sensed while the beam is decentered from the center in different directions indicated by the colored arrows.
  The different lines in image \textbf{(b)} correspond to these different directions of this tip-tilt motion, ranging from $0^{\circ}$ (blue) to $60^{\circ}$ (green) in steps of $6^{\circ}$ and in reference to the sensing \ac*{MMF}.
  }
	\label{fig:coupling_60deg}
\end{figure}

\begin{figure}
	\begin{center}
		\begin{tabular}{c c}
			\subfloat[]{
        \includegraphics[height=6cm,trim={0 0 0 300},clip] {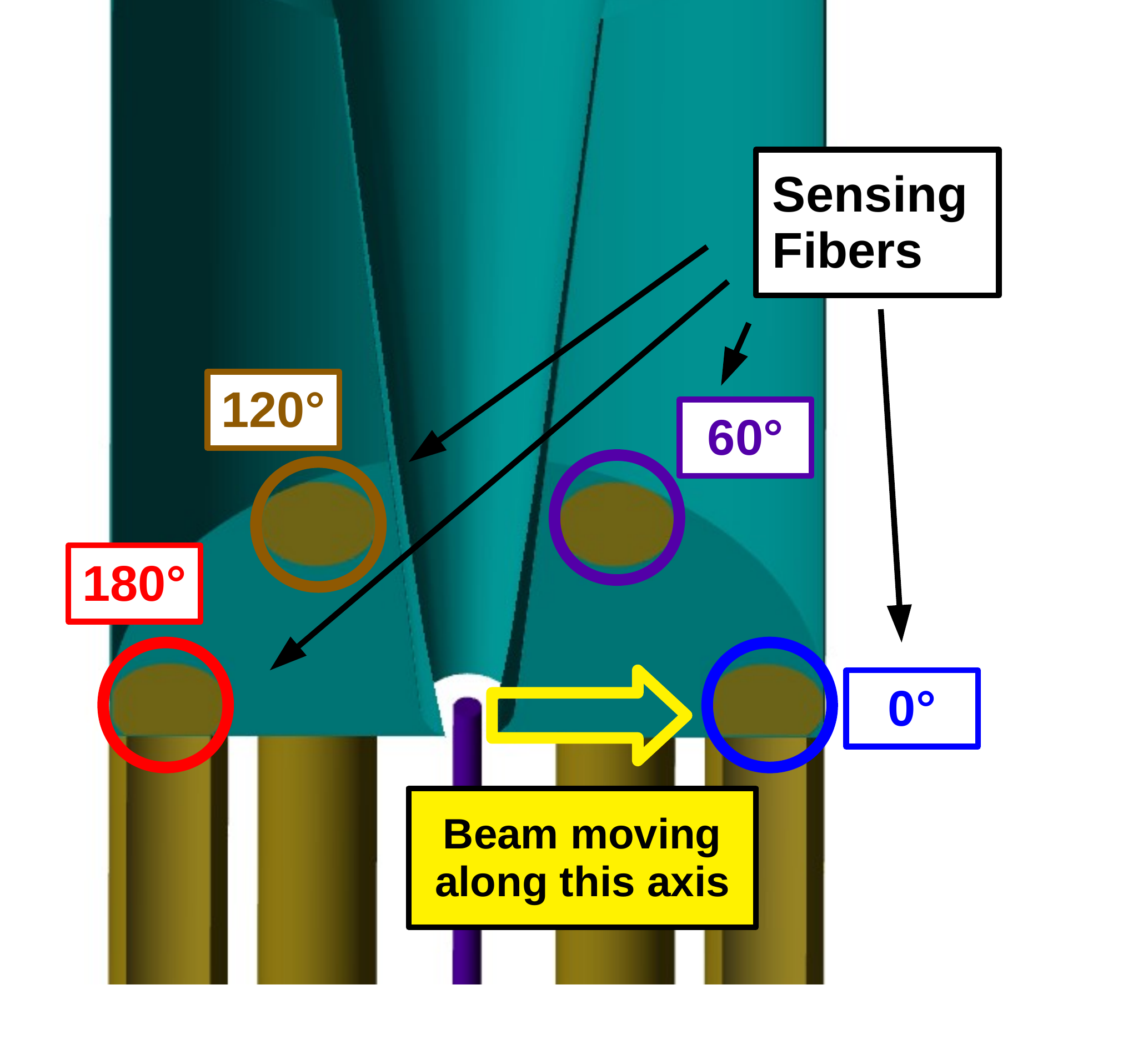}
        \label{fig:coupling_360deg_model}
        } &
			\subfloat[]{
        \includegraphics[height=6cm] {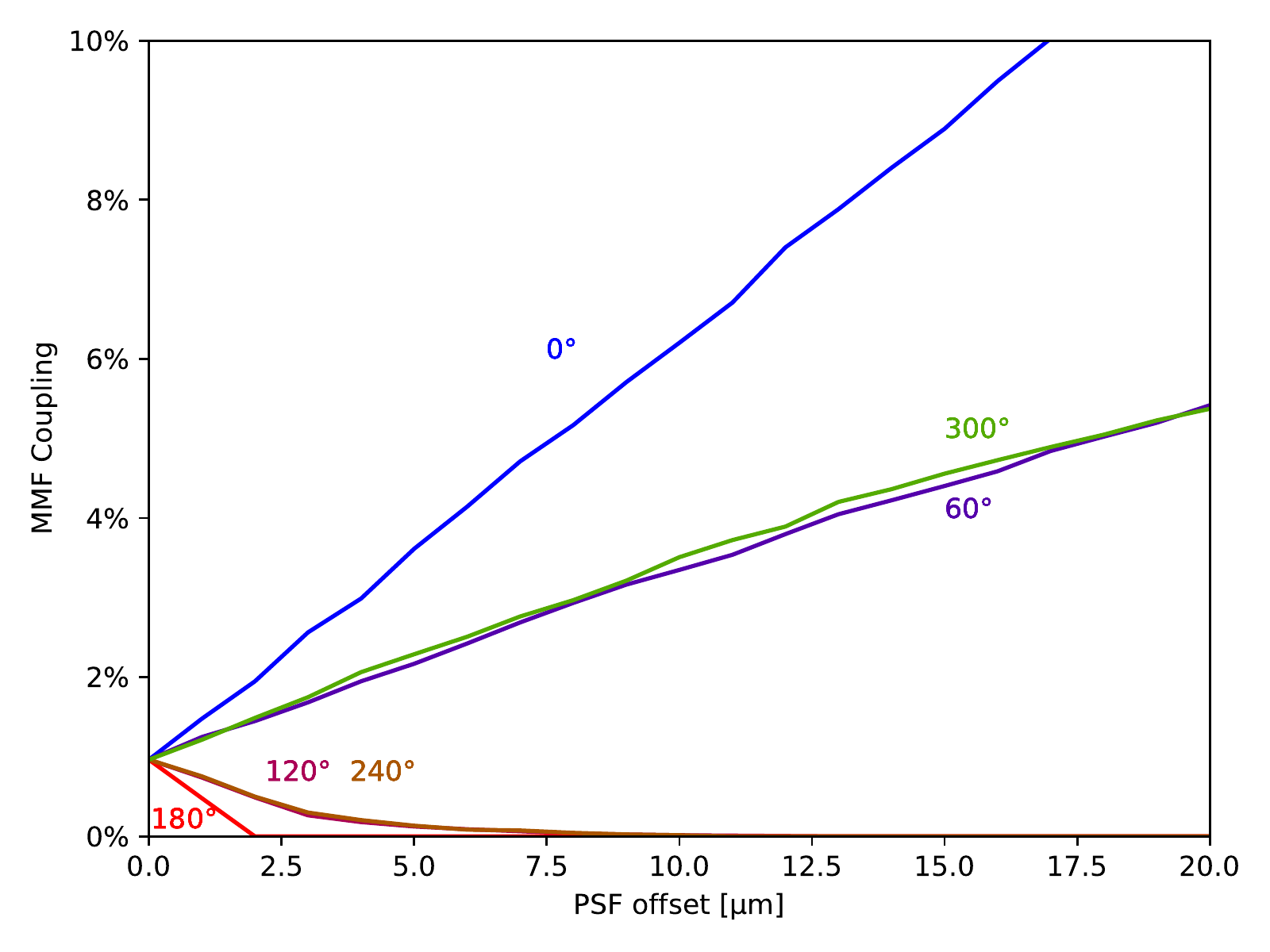}
        \label{fig:coupling_360deg_plot}
        }
		\end{tabular}
	\end{center}
	\caption{
  Sensing signal for each \ac*{MMF} as the incoming beam moves from an aligned position outwards as illustrated in image \textbf{(a)} (two fibers are not shown).  Image \textbf{(b)} shows the signal as function of the misalignment.
  The signal of the three fibers in direction of the beam increases linearly while the signal on the opposite fibers decreases linearly.
  }
	\label{fig:coupling_360deg}
\end{figure}

To examine the performance if the beam is not aligned along an axis of a sensing \ac{MMF}, the sensing signal is modeled for such misaligned beams.
This is plotted in Fig.~\ref{fig:coupling_60deg_plot} showing a misaligned beam scanning along eleven different axes ranging from alignment to the \ac{MMF} axis to a rotation of $60^{\circ}$ corresponding to the next sensing fiber.
This is illustrated on the model in Fig.~\ref{fig:coupling_60deg_model}.
The linear response that was already observable in Fig.~\ref{fig:MLA_coupling} is seen to extend up to a misalignment of $70~\mum$.
This is valid for both the aligned beam ($0^{\circ}$, blue) and a beam centered exactly between two \acp{MMF} ($30^{\circ}$, red).
This very wide dynamical range enables the usage for not only real-time tip-tilt sensing, but also for initial fiber and target alignment where a large dynamical range is favorable.

A similar analysis was performed to evaluate the sensing response of all \acp{MMF} for a misaligned beam.
This is shown in Fig.~\ref{fig:coupling_360deg_plot} with a schematic in Fig.~\ref{fig:coupling_360deg_model} illustrating the procedure. As the beam moves towards the rightmost fiber (blue), coupling is measured for all six sensing fibers (two are not shown in the illustration) numbered by the angular position in respect to the rightmost fiber.
All signals show an initial linear response with different slopes allowing reconstruction of the actual beam position.
As there is an initial response in all sensing fibers, some errors signals from other aberrations or detector noise can be filtered.

\subsubsection{Central aperture}
\label{sec:MLA_aperture}

\begin{figure}
  \centering
  \includegraphics[width=0.6\textwidth]{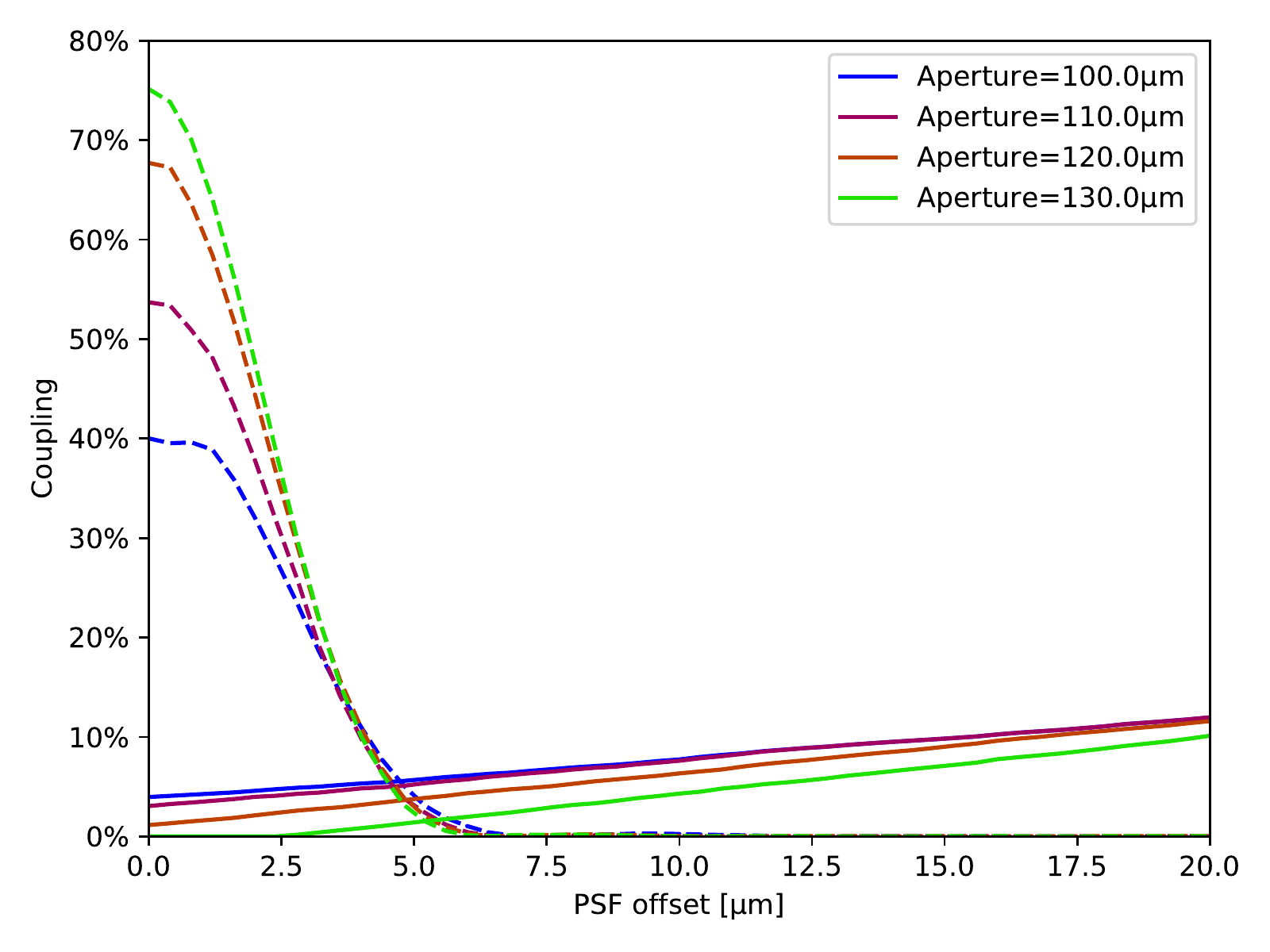}
  \caption{
  Tip-tilt sensor performance for different central apertures ranging from $100~\mum$ (blue) to $130~\mum$ (green) in diameter. Dashed lines correspond to central \ac{SMF} coupling and solid lines to sensing \ac{MMF} coupling.
  For larger central apertures, the maximum coupling efficiency into the \ac{SMF} increases but the sensing signal decreases. This trade-off between maximum coupling and sensing sensitivity needs to be evaluated and chosen to fit the requirements of the system.
  }
  \label{fig:MLA_aperture}
\end{figure}

The \ac{MLA} was designed such that $\sim70\%$ of the total intensity is coupled into the central \ac{SMF} if the beam is aligned.
This is less than the theoretical maximum of $\sim80\%$ \cite{Shaklan1988} and is due to the fact that the central void part of the lens, i.e. the unobscured area in front of the \ac{SMF}, is smaller than necessary for maximum coupling efficiency.
This way, a sensing signal in the \ac{MMF} can be provided even if the beam is aligned, thus providing stability as sensing can be performed even for a very small beam offset.
Fig.~\ref{fig:MLA_aperture} shows a range of possible choices for the diameter of the central aperture between $100~\mum$ (blue) to $130~\mum$ (green).
If the systems requirements demand a higher sensing signal or on the contrary a higher \ac{SMF} coupling, the design may be adjusted accordingly. For a reasonable trade-off, we have decided on a central aperture of $\sim120~\mum$ that yields both good coupling ($\sim70\%$) and a good sensing signal even for an aligned beam ($\sim1\%$).

\subsection{Tip-tilt correction}
\label{sec:correction}

The \ac{MLA} tip-tilt sensor described is this section provides a signal indicating that the incoming beam is misaligned.
This sensor signal will then be transformed into a correction signal which feeds a tip-tilt mirror within the iLocater front-end prototype.
The sensor and the mirror will then be able to correct in real-time and closed-loop.
To read out the sensing fibers, a fast, low-noise photo-detector can be used such that the correction frequency is governed by the photon count of the science target.

As mentioned in Sec.~\ref{sec:MLA_axis}, there is a response in all six sensing fibers if the beam is decentered.
This enables the development of an interaction matrix which takes into account possible error signals that occur due to other aberrations than tip-tilt as well as background noise.

\subsection{Manufacturing}
\label{sec:manufacturing}

Current approaches for fabricating individually lensed fibers rely on grinding \cite{Yeh:2005}, etching \cite{Eisenstein:1982} or melting techniques \cite{Presby:1990}. These require complex fabrication processes that are not well suited for \acp{MCF} or compact fiber bundles. In order to fabricate these, \acp{MLA} can be glued onto the fibers. These lenses, however, must be aligned with the fibers in three or six degrees of freedom, thereby considerably complicating the assembly, particularly for \ac{SM} fibers, where alignment tolerances are very stringent.

Recently it has been demonstrated, that in-situ fabricated beam-shapers allow ultra low-loss coupling for a variety of application\cite{Dietrich2018} including astrophotonics\cite{Dietrich2017}.
This technology makes use of 3D-lithography by two-photon polymerization of a commercial IP-resist by \textit{nanoscribe}\cite{nanoscribe}.
This solves several problems: Due to the flexibility of 3D-lithography lens designs can be adapted rapidly to any requirements of the optimal system, ensuring best possible coupling. This is particularly useful if the cores of the \ac{MCF} or fiber bundle are not regularly spaced. Additionally, the in-situ fabrication of lenses circumvents tedious alignment steps. Additionally, free-form lenses can be formed in shapes that are difficult to manufacture using conventional methods. They do not require any adhesives that may decrease both short-term and long-term stability.

\section{KOOL: AO system testbed}
\label{sec:KOOL}

\begin{figure}
  \centering
  \includegraphics[width=\textwidth]{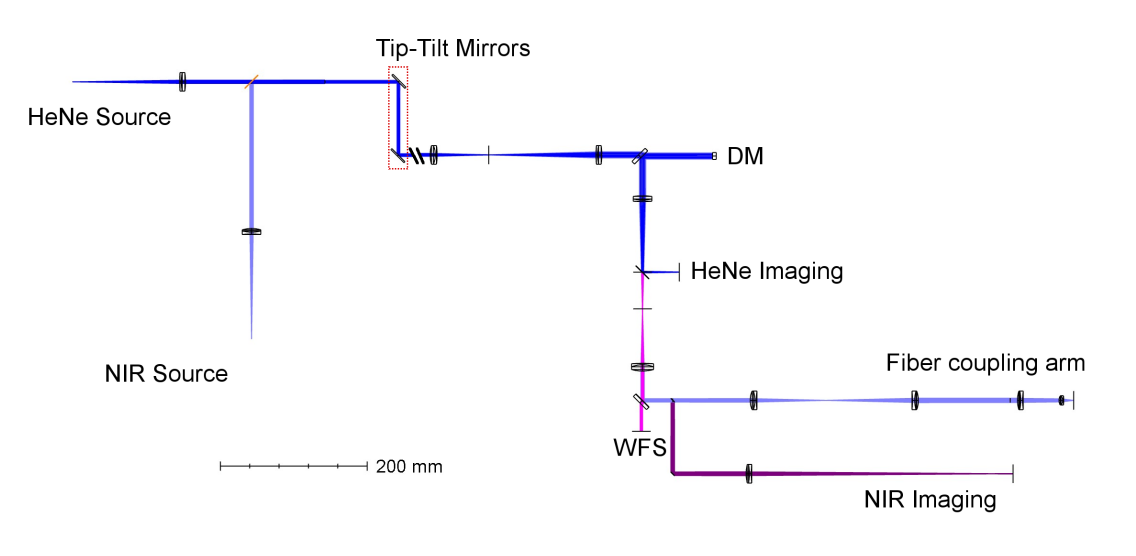}
  \caption{
    Design layout of the \acf*{KOOL}.
    This \acf*{AO} testbed allows simulation and correction of \ac*{LBT} vibrations with two tip-tilt mirrors and higher order aberrations with an ALPAO \acf*{DM}.
    A HeNe Source (632 nm) is used in the main setup including the \acf*{WFS} (purple) and an imaging arm (dark blue). A \acf*{NIR} ($1.31~\mum$) source feeds both a fiber coupling arm (light blue) and a separate imaging arm (violet).
    Both sources are \ac*{SMF}-fed.
  }
  \label{fig:KOOL_design}
\end{figure}

We have developed an optical tested for performing realistic tests for vibration correction and single-mode fiber coupling in large telescopes, which we call \acf{KOOL}.
This is a collaboration between the \ac{MPIA} in Heidelberg, Germany, \ac{ISYS} in Stuttgart, Germany and the Landessternwarte Heidelberg (LSW, part of Zentrum für Astronomie der Universität Heidelberg) in Heidelberg, Germany.

Fig.~\ref{fig:KOOL_design} shows the schematics of the setup.
This testbed is separated into two sections.
Beam manipulation optics make use of several opto-mechanical components to generate and manipulate a test beam that will simulate an incoming telescope beam.
A first tip-tilt mirror can introduce vibrations up to 50~Hz into the optical system, which can be similar to the \ac{PSD} of the LBT. Furthermore, the mirror is equipped with several accelerometers to test the disturbance feed-forward control in real time \cite{Gluck2017}. Based on the accelerometers the low order aberrations as piston, tip and tilt can be estimated on-line by a linear filter. A second tip-tilt mirror as well as the DM can be used for the compensation of the vibrations. The frequency range of the tip-tilt correction is up to 1~kHz.
A phase screen (not shown) is also available and can introduce atmospheric aberrations with variable speed.
The \ac{DM} is able to either correct for those aberrations in closed-loop or it can itself introduce desired aberrations to simulate a certain environment or a known telescope \ac{PSF}.
For both closed-loop operation, as well as for wavefront quality control, this setup includes a \ac{WFS} and a camera for imaging the \ac{PSF}.
All beam manipulation and closed loop operation is performed with a \ac{SMF}-fed HeNe source (632nm).

A second, \ac{NIR} source ($1.31~\mum$) is also fed into the system with a dichroic mirror.
It passes through the \ac{AO} system and it is then again reflected by a dichroic mirror, entering the fiber coupling section.
This section consists of a fiber coupling arm to test and characterize \ac{SMF} coupling and the final \ac{MLA} tip-tilt sensor.
By using a beam splitter, the \ac{NIR} source is also imaged onto a camera to compare fiber coupling results to the \ac{PSF}.

\section{Discussion}
\label{sec:discussion}

Other options to sense tip-tilt aberrations include a pinhole mirror near the focal plane to reflect any misaligned light, a beam splitter or dichroic mirror to image the \ac{PSF} on a \ac{CCD} or quad-cell detector\cite{Esposito1997}, the \ac{AO} system of the telescope, or an accelerometer based feed-forward control\cite{Gluck2017}, where the accelerometers are mounted at the telescope mirros and the reconstructed low order aberrations are used for compensation in a disturbance feed-forward combined with the AO system.
Yet, the \ac{MLA} tip-tilt sensor design introduced in this work yields many advantages but also some challenges.
To compensate for difficulties it is also possible to combine different approaches. Such a system could combine a feed-forward system for rough tip-tilt compensation and the \ac{MLA} tip-tilt sensor for closed-loop high precision correction.

\begin{description}
  \item[Throughput]
    The \ac{MLA} tip-tilt mirror will couple almost all light into the \ac{SMF} if the beam is aligned. This can be increased in trade-off for sensitivity (see Sec.~\ref{sec:MLA_aperture}).
    When imaging the \ac{PSF} using a beam splitter, some light needs to to be diverted and is therefore not available for coupling. This is not the case when using a dichroic mirror as sensing is done in a different wavelength range.
    There will also be no losses when using either the telescope \ac{AO} system as no additional light is diverted or when using an accelerometer based feed-forward system as no light is detected all together.
    When using a pinhole mirror, the alignment precision of the pinhole to the \ac{SMF} will govern its throughput and can lead to large losses.
  \item[Vignetting]
    Light may be vignetted by a pinhole in front of the focal plane causing reduced coupling. This is especially worth considering as the pinhole mirror needs to reflect the beam at an angle leading to an elliptical aperture.
    The \ac{MLA} tip-tilt sensor also suffers from vignetting as the central unobscured area in front of the \ac{SMF} acts as an aperture. However, this can be modeled reliably as in-situ printing of the \ac{MLA} assures good alignment and its circular shape assures symmetry.
    All \ac{PSF} imaging, \ac{AO} correction and feed-forward control will not lead to any vignetting.
  \item[Range]
    The dynamical range is rather large on all afore mentioned options. An exception is the \ac{AO} system of the telescope as a large tip-tilt error in the wavefront  may already be outside the dynamical range of the \ac{WFS}.
    The \ac{MLA} brings the additional advantage that the sensor provides a linear signal for a very large range. Due to these properties the sensor can also be used for initial alignment on the target.
  \item[Sensitivity]
    The \ac{MLA} tip-tilt sensor ensures excellent sensitivity, though some of it is penalized by a lower maximum fiber coupling efficiency. Additionally, as the sensing \acp{MMF} can be read out individually, the possibility to use low-noise read out electronics can further increase sensitivity.
    As a pinhole in front of the \ac{SMF} practically needs to be larger than the fiber to assure high throughput, sensing of slight misalignment will not be challenging.
    Also, when imaging either the pinhole mirror or the \ac{PSF} directly on a \ac{CCD}, readout noise will limit the sensitivity.
    Using a feed-forward system that makes use of an accelerometer will in general be quite sensitive to small amplitude vibrations.
    The disturbance feed-forward system mainly considers the vibrations in the telescope path. However, vibrations are also introduced in the instrument path. To achieve optimal coupling an additional sensor as the \ac{MLA} is needed in a closed-loop system for high precision coupling.
  \item[Speed]
    Again, as readout electronics can be chosen more freely, fast photo-diodes can be used for the \ac{MLA} tip-tilt sensor in combination with a simple tip-tilt correction algorithm allowing correction frequencies that are only limited by photon count.
    All sensors that rely on sensing with a \ac{CCD} or a quad-cell detector are limited by its readout and signal processing sampling.
    On the other side, a feed-forward system can also work quite fast as the tip-tilt detection is independent of the photon collection. Therefore the speed limitations only depend on the mechanical dynamics of the sensor and the electronics.
  \item[Chromaticity]
    Even though tip-tilt aberrations are known to be achromatic,
    it is worth while mentioning that imaging the \ac{PSF} with a dichroic mirror will lead to tip-tilt sensing in a different wavelength range.
    An accelerometer based approached is also wavelength independent leading to possible chromatic effects.
    All other tip-tilt sensor options detect in the working wavelength range.
  \item[Non-common path aberrations]
    Tip-tilt vibrations that occur between the sensing mechanism and the focal plane can cause most of the fiber coupling inefficiency. This is mostly true for the \ac{AO} system \ac{WFS} and any accelerometer that will potentially be much further upstream. Separate imaging of the \ac{PSF} can also be affected by non-common path aberration, strongly depending on where it is integrated into the optical system.
    As sensing is done right before the focal plane for both the pinhole mirror technique and \ac{MLA} tip-tilt sensor, all vibrations throughout the system are detected and can be corrected for.
    This also leads to the exciting possibility to not only use the \ac{MLA} for real-time tip-tilt sensing but also to correct for static higher order aberrations. As sensing is done near the focal plane and in the working wavelength range, overall coupling efficiency can be optimized by correcting with the \ac{DM}. Furthermore, as there is data from one central \ac{SMF} and six surrounding sensing fibers available, the optimization algorithm can make use of more data than e.g. a quad-cell detector.
  \item[Size and complexity]
    This is one of the main advantages of the \ac{MLA} tip-tilt sensor. While all other closed-loop systems depend on major modifications of the optical setup, the \ac{MLA} can be integrated easily without changing the optical train. For this, the fiber bundle with the \ac{MLA} is placed at the focal plane and readout electronics can be mounted remotely.
    This compact design reduces complexity and cost of the implementation into the system. On the other hand, design and manufacturing of the fiber bundle and of the \ac{MLA} itself are more complex but also open up many possibilities as the free-form lenses allow adjustments to a wide range of optical requirements and performance goals.
\end{description}

\subsection{Comparison to Prototype}
\label{prototype}

\begin{figure}
	\begin{center}
		\begin{tabular}{c c}
			\subfloat[]{
        \includegraphics[height=10cm] {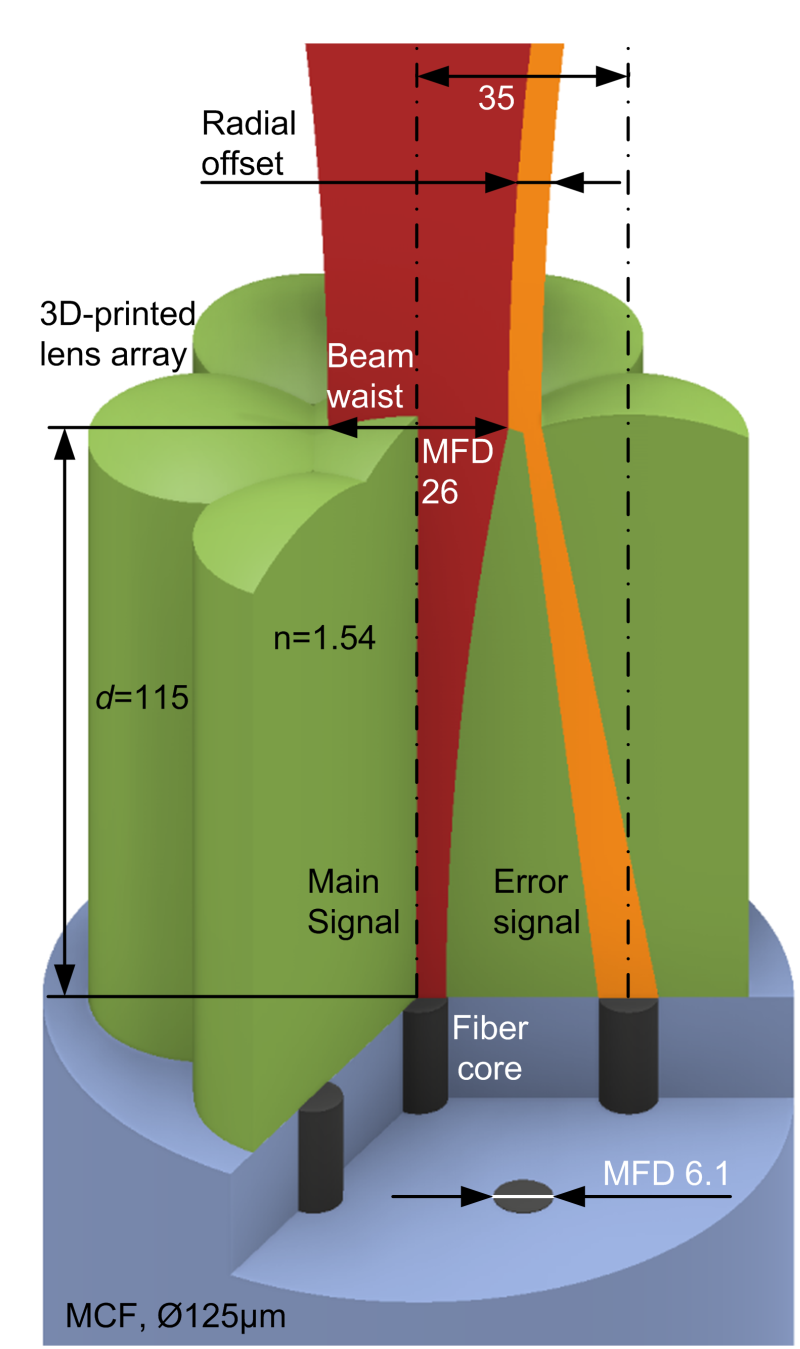}
        \label{fig:prototype-model}
        } &
			\subfloat[]{
        \includegraphics[height=5cm] {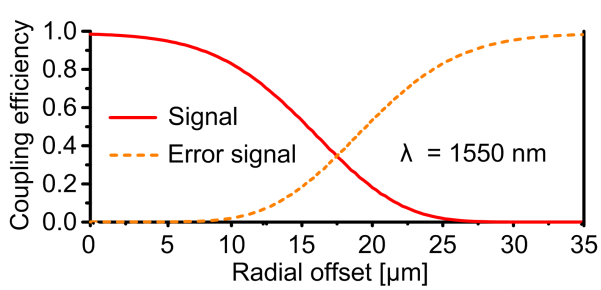}
        \label{fig:prototype-coupling}
        }
		\end{tabular}
	\end{center}
	\caption{
  Prototype \ac{MLA} (green) printed on top of a \acf{MCF} (blue) consisting of seven \acfp{SMF} (black) as introduced by Ref.~\citenum{Dietrich2017}.
  Image \textbf{(a)} shows a 3D~model of the design.
  Image \textbf{(b)} shows the \acf{SMF} coupling efficiency normailized to the maximum coupling efficiency for both the the central fiber (red line) and the outer sensing fiber (dashed orange line). Note the different sensing response in comparison to the design introduced in this work as plotted in Fig.~\ref{fig:MLA_coupling}.
  Images modified and reprinted from Ref.~\citenum{Dietrich2017}.
  }
	\label{fig:prototype}
\end{figure}

An initial design and prototype was introduced in 2017\cite{Dietrich2017} (3D~model shown in Fig.~\ref{fig:prototype-model}).
The design in this work is based upon this prototype but incorporates a number of major changes.
First of all, the prototype design also refracts the incoming beam when it is aligned to the optical axis and focuses it onto the \ac{SMF}.
The new design from this work offers two major improvements in that regard. By allowing the aligned beam to pass to the fiber coupling plane without refraction, the design from this work offers two improvements. Firstly, the optical path does not need to be modified for the \ac{MLA} tip-tilt sensor to be integrated into the system.
Secondly, this design guarantees maximum performance if the beam is aligned as the optical system is designed for maximum coupling efficiency.
Therefore, possible error sources such as reflection, absorption, limited surface quality and chromaticity of the lens material can be disregarded.

Furthermore, the prototype makes use of \acp{SMF} for sensing, leading to a sensing signal that is very similar to that of the central \ac{SMF} that leads to the instrument. This is plotted in Fig.~\ref{fig:prototype-coupling}.
The design introduced in this work, on the other hand, is able to create a linear response because the usage of \acp{MMF} allow more tolerances and enable a more efficient fiber coupling (compare to Fig.~\ref{fig:MLA_coupling}).
As using \acp{SMF} do not offer any advantage for the sensor readout, the only disadvantage of using \acp{MMF} is an increased coupling of spatially separated objects or other aberrations than tip-tilt.

\section{Conclusion}
\label{sec:conclusion}
\acresetall

In this work we have introduced a preliminary design for a tip-tilt sensor with integrated \ac{SMF} coupling that is optimized to be used with the prototype front-end of the \ac{SM} spectrograph iLocater.
This design can be integrated into the existing fiber coupling optics without any modifications.
When sensing tip-tilt motion of the incoming beam, modeled performance yields a linear response, which simplifies signal processing correction algorithms.
The device can be modified to fit system requirements and performance goals.

We have discussed the advantages of this sensor when compared to conventional tip-tilt sensing options.
This includes the compactness of the device, the capability to integrate it into existing optical systems easily, the sensing at the focal plane to avoid \ac{NCP} vibrations and a higher sensitivity and sampling frequency as detectors can be chosen much more freely.
Due to its wide dynamical range, this design can be used for initial fiber and target alignment. Furthermore, it can be used to feed \ac{NCP} aberration optimization algorithms.

We have also introduced the adaptive optics (AO) testbed KOOL which can be used to introduce and correct \ac{LBT} vibrations and higher order aberrations.
We will use this testbed to test, characterize and optimize this device.

We are currently in the final design stages and will be manufacturing the final device soon. This will be tested at \ac{KOOL} and then be integrated and tested at the \ac{LBT}.

\acknowledgments

This work was supported by the Deutsche Forschungsgemeinschaft (DFG) through project 326946494, 'Novel Astronomical Instrumentation through photonic Reformatting'.

This publication makes use of data generated at the Königstuhl Observatory Opto-mechatronics Laboratory (short: KOOL) which is run at the Max-Planck-Institute for Astronomy (MPIA, PI Jörg-Uwe Pott, \url{jpott@mpia.de}) in Heidelberg, Germany. KOOL is a joint project of the MPIA, the Landessternwarte Königstuhl (LSW, Univ. Heidelberg, Co-I Philipp Hottinger), and the Institute for System Dynamics (ISYS, Univ. Stuttgart, Co-I Martin Glück). KOOL is partly supported by the German Federal Ministry of Education and Research (BMBF) via individual project grants.

\bibliography{phd}
\bibliographystyle{spiebib} 

\end{document}